# How Can Video Generative AI Transform K-12 Education? Examining Teachers' Perspectives through TPACK and TAM

Unggi Lee[1], Yeil Jeong[†2], Seungha Kim[*3], Yoorim Son[*4], Gyuri Byun[3],

Hyeoncheol Kim[6], & Cheolil Lim[3]

Enuma, Inc., AI Engineering Team, Seoul, South Korea[1]

Intercollegiate Initiative for Talent Development, Seoul National University, Seoul, South Korea[2]

Seoul National University, Department of Education, Seoul, South Korea[3]

Pungnap Middle School, Seoul Metropolitan Office of Education, Seoul, South Korea[4]

Department of Computer Science and Engineering, Korea University[5]

codingchild@korea.ac.kr / yell001@snu.ac.kr / seungha@snu.ac.kr /

thsvlzk@naver.com / gyuribyun97@gmail.com / harrykim@korea.ac.kr / chlim@snu.ac.kr

*Equal contribute, †Corresponding author

**Abstract**

The rapid advancement of generative AI technology, particularly video generative AI (Video GenAI), has opened new possibilities for K-12 education by enabling the creation of dynamic, customized, and high-quality visual content. Despite its potential, there is limited research on how this emerging technology can be effectively integrated into educational practices. This study explores the perspectives of leading K-12 teachers on the educational applications of Video GenAI, using the TPACK (Technological Pedagogical Content Knowledge) and TAM (Technology Acceptance Model) frameworks as analytical lenses. Through interviews and hands-on experimentation with video generation tools, the research identifies opportunities for enhancing teaching strategies, fostering student engagement, and supporting authentic task design. It also highlights challenges such as technical limitations, ethical considerations, and the need for institutional support. The findings provide actionable insights into how Video GenAI can transform teaching and learning, offering practical implications for policy, teacher training, and the future development of educational technology.

*Keywords: Video Generative AI, K-12 Education, TPACK, TAM, Teacher Perspectives*

**Introduction**

Visual media has been a crucial teaching and learning tool in education, enabling the visualization of complex scientific phenomena, historical event reenactments, and the concretization of abstract concepts (Brame, 2017; Noetel et al., 2021; Woolfitt, 2015). In particular, video content, which incorporates visual elements, movement, and temporality, has proven effective in enhancing student understanding and engagement (Brame, 2017; Gaudin & Chaliès, 2015; Zhang et al., 2006). However, creating or acquiring videos suitable for curriculum and learner levels has been challenging in educational settings due to time, cost, and technical constraints. This challenge is particularly acute for content that is difficult to film directly, such as scientific experiments or historical events (Norton & Hathaway, 2010; Premlatha & Geetha, 2015).

Recent advances in generative AI (GenAI) technology present potential solutions to these limitations. Notably, Sora, released by OpenAI in early 2024 (OpenAI, 2024b), is a Video GenAI model capable of creating realistic, high-quality videos from natural language prompts. Unlike existing text or image GenAI, this technology enables easy creation of content incorporating movement and temporality, drawing significant attention from the educational community (Dao et al., 2021; Yu et al., 2024). Teachers can now potentially create customized videos that align with curriculum requirements and student needs, promising innovative changes in teaching and learning methods.

However, research on the educational applications of Video GenAI remains in its early stages. Previous studies on GenAI have primarily focused on text generation models like ChatGPT (Adeshola & Adepoju, 2024) or image generation models like Stable Diffusion (Lee et al., 2023) and DALL-E (OpenAI, 2022a; Ramesh et al., 2021). These studies have reported that

GenAI can provide personalized learning experiences, enhance teacher efficiency, and enable new forms of assessment (Adeshola & Adepoju, 2024; Lo, 2023; Opara et al., 2023). Yet, systematic exploration of the unique characteristics and educational potential of Video GenAI has not been conducted.

Notably, there is limited research examining teachers' perceptions and expectations of how this technology could be implemented in actual educational settings. Teachers play a pivotal role in adopting new technologies to the field of education, as they are the primary decision-makers responsible for budget allocation and the practical implementation of these tools (Almuhanna, 2024; Jhurree, 2005; Viberg et al., 2024). They serve as hands-on practitioners who select AI programs for the classroom use, while simultaneously acting as critical evaluators who assess the potential benefits and limitations of these technologies before introducing them to their students (Alwaqdani, 2024; Shi et al., 2024; Zhai, 2024).

Gaining teachers' approval and persuading them of the utility of such innovations are crucial steps in enhancing the overall quality of education technology (Granić, 2022; Mazman Akar, 2019; Stumbrienė et al., 2024). Research has emphasized the importance of teacher agency in successful educational innovation (Brown et al., 2023; Cong-Lem, 2024; Priestley et al., 2012) and highlighted how teacher perceptions and competencies are decisive in implementing new digital technologies (Mulyadi et a., 2023; Wozney et al., 2006). The willingness of users to continue utilizing technological tools or platforms largely depends on their perceived experiences and value (Berbar, 2023; Davis, 1989; Deng et al., 2010; Zhang et al., 2022). Consequently, investigating teachers' initial impressions and usage experiences with Video GenAI is essential to identify key factors that influence their acceptance and inform strategies for effective integration into educational practices.

This study aims to analyze leading teachers' perceptions and expectations of educational applications of Video GenAI. Specifically, it explores how this technology can enable meaningful educational tasks and what factors need to be considered in their implementation. The significance of this study lies in three aspects. First, it examines the potential educational applications of Video GenAI from practicing teachers' perspectives, leading to practical and feasible implementation strategies. Second, by focusing on authentic task design, it provides concrete insights into how this technology can translate into meaningful learning experiences. Third, by identifying barriers to technology adoption and necessary support measures, it offers implications for future policy directions.

The research questions for this study are:

1. What are teachers' first impressions and usage experiences of Video GenAIs?

2. What potential do Video GenAIs have for educational applications?

3. What are the limitations of Video GenAIs in educational applications, and what are the ways to improve them and future development directions?

<center>**Literature Review**</center>

**Generative AI in Education**

The rise of AI has significantly transformed various sectors, including education. Among its developments, GenAI, such as OpenAI's ChatGPT (OpenAI, 2022b), has gained widespread attention for its potential to personalize learning, enhance instructional design, and foster creative problem-solving (Alasadi & Baiz, 2023; Khan et al., 2023; Qadir, 2022). Positioned as a transformative force, GenAI's applications in education span both learner-centered customization and teacher-driven assessment.

First, GenAI enables personalized and interactive learning experiences. Unlike traditional one-way instruction, GenAI can engage students in authentic scenarios, adapting content based on individual learning styles and needs (Su & Yang, 2023). This data-driven adaptability enhances engagement while offering insights into students' strengths and areas for improvement (Ayeni et al., 2024). Second, GenAI supports automated assessment and feedback, reducing teacher workload while increasing efficiency and accuracy in evaluation (González-Calatayd et al., 2021). Despite limitations such as scope and ethical concerns, AI-driven assessment tools provide real-time, personalized feedback, benefiting students—particularly those seeking adaptive, metacognitive learning experiences (Owan et al., 2023; Shetye, 2024).

However, integrating GenAI into education also raises critical ethical and pedagogical concerns. First, issues such as data privacy and algorithmic bias remain unresolved, as GenAI systems may inadvertently reinforce discrimination or expose sensitive information (Halaweh, 2023). Second, concerns about cognitive dependency suggest that excessive reliance on GenAI could undermine students' problem-solving and critical thinking skills (Iskender, 2023;

Sánchez-Ruiz et al., 2023). Finally, GenAI's tendency to generate misinformation presents risks to knowledge acquisition, requiring students to develop stronger self-regulation and verification skills (Ahmad et al., 2023). Given these challenges, continued scrutiny is necessary to establish ethical guidelines that align AI applications with sound educational objectives.

As AI evolves, its role in education is shifting beyond text-based applications toward multimodal GenAI, including Video GenAI (Lee et al., 2024a; Mittal et al., 2024). Despite its potential to enhance engagement and facilitate dynamic learning experiences, research on educators' perceptions and practical integration strategies remains limited (Lee et al., 2024b; Leiker et al., 2023). Addressing this gap, this study explores teachers' perspectives on Video GenAI, examining both its educational applications and the institutional, pedagogical, and ethical considerations that shape its adoption.

**TPACK and TAM for AI**

The integration of AI in education has elicited varied responses from teachers, influenced by digital competence, pedagogical beliefs, and ethical concerns (Galindo-Domínguez et al., 2024; Wang et al., 2021). While some educators view AI as a valuable tool despite limited experience, others hesitate due to concerns about plagiarism, misinformation, and pedagogical misalignment (Iqbal et al., 2022; Yue et al., 2024). As AI technologies become more prevalent, understanding how teachers perceive and integrate them is crucial (Cukurova et al., 2023; Nazaretsky et al., 2022). To analyze these dynamics, established theoretical frameworks such as Technological Pedagogical Content Knowledge (TPACK) and the Technology Acceptance Model (TAM) offer critical insights.

TPACK explains technology integration by examining the interplay between content knowledge (CK), pedagogical knowledge (PK), and technological knowledge (TK) (Koehler & Mishra, 2009). This model highlights how teachers must align AI with instructional methods and subject matter, considering not only how AI works (TK) but also how it supports pedagogical strategies (TPK) and content delivery (TCK). TAM, on the other hand, focuses on Perceived Usefulness (PU) and Perceived Ease of Use (PEU) as key determinants of technology adoption (Davis et al., 1989). Teachers are more likely to integrate AI if they perceive it as beneficial and easy to use (Marangunić & Granić, 2015), with additional factors such as trust and ethical considerations influencing adoption (Iqbal et al., 2022).

The emergence of Video GenAI presents new considerations for these frameworks. Unlike traditional educational technologies, Video GenAI enables dynamic, multimodal content creation, raising questions about instructional alignment, content validity, and ethical use (Mishra et al., 2023). TPACK suggests that teachers need to navigate both technological and pedagogical complexities, ensuring that AI-generated videos align with curriculum goals. Meanwhile, TAM indicates that perceptions of accessibility and instructional value will shape teachers' willingness to adopt this technology (Davis, 1989).

This study applies TPACK and TAM to investigate how teachers perceive and adopt Video GenAI as an instructional tool. Through qualitative interviews and hands-on experimentation, we explore the factors that influence teachers' willingness to integrate Video GenAI, the pedagogical strategies they develop, and the institutional support required for effective implementation. By shifting the focus from a tool-centric perspective to a context-driven analysis of AI in education, this study offers insights into how Video GenAI can be meaningfully integrated into K-12 learning environments.

**Video Generative AI**

Since OpenAI introduced Sora in 2024 (OpenAI, 2024; Liu et al., 2024b), Video GenAI has gained increasing attention. Efforts to enhance large-scale video generation models have led to significant advancements, particularly through training on vast datasets (Bar-Tal et al., 2024; Hong et al., 2023; Kondratyuk et al., 2023; Villegas et al., 2023).

One of the most influential technologies in this domain is the Diffusion Transformer (DiT) (Gao et al., 2023; Peebles & Xie, 2023), which has become a cornerstone for state-of-the-art video generation. DiT integrates diffusion models' iterative denoising (Ho et al., 2020) with Transformers' scalability and attention mechanisms (Vaswani et al., 2017), allowing it to effectively model temporal and spatial dependencies (Liu et al., 2024b). This approach enhances visual coherence across frames, enabling the creation of high-resolution, dynamic videos from textual prompts. Models leveraging these advancements—including OpenAI Sora, Kling AI's Kling, Luma Dream Machine, Runway Gen-3, and Google DeepMind's Veo 2—have set new benchmarks for text-to-video synthesis (OpenAI, 2024; Kling AI, 2024; Luma AI, 2024; Runway AI, 2024; Google DeepMind, 2024).

Video GenAI's potential in education lies in its ability to democratize media production, allowing students to create and engage with personalized, visually rich content (Pellas et al., 2023). Tools like HeyGen and Synthesia streamline video creation, enhancing accessibility while reducing teachers' workloads (Dao et al., 2021; Yu et al., 2024). However, challenges remain, including concerns about technological limitations, misinformation, and ethical risks such as bias in AI-generated content (Adetayo et al., 2024; Shu et al., 2021).

Despite these challenges, existing research has primarily relied on smaller-scale models, limiting their effectiveness in educational contexts. To address this gap, our study investigates the pedagogical applications of cutting-edge models like the DiT. By exploring teachers' perspectives through the TPACK and TAM frameworks, we assess the practical utility of Video GenAI in K-12 learning environments and identify strategies for effective integration.

## Method

### Participants and Procedure

This study employed a qualitative research approach to investigate teachers' perceptions and expectations of Video GenAI in educational contexts. Ten teachers participated in this study: eight elementary school teachers, one high school teacher, and one data analyst with elementary teaching experience (See **Table 1**). Participants were purposefully selected based on their experience with educational technology and AI implementation in teaching, as the study aimed to explore potential best practices and challenges from teachers with higher-than-average experience and interest in GenAI-related tools, rather than to generalize findings to all teachers. Additionally, to minimize the impact of initial attitudes toward AI on the study's findings, participants were selected based on their prior interest and experience in using AI tools.

**Table 1**

*The Information of the Participants*

| Name | Occupation | Academic Degree | Major | Career (years) |
|------|-----------|-----------------|-------|----------------|
| T1 | Elementary teacher | Ph.D. | Elementary education, computing education | 9 |
| T2 | Elementary teacher | Ph.D. student | Elementary education, Educational technology | 10 |
| T3 | Elementary teacher | M.Ed. | Elementary education, gifted education | 16 |
| T4 | Elementary teacher | M.Ed. | Elementary education, AI education | 8 |
| T5 | Elementary teacher | M.Ed. | Elementary education, AI education | 9 |
| T6 | Elementary teacher | M.Ed. | Elementary education, computing education | 13 |
| T7 | Elementary teacher | Bachelor | Elementary education | 9 |
| T8 | Elementary teacher | Bachelor | Elementary education | 8 |
| T9 | High school English teacher | M.A. | English education | 11 |
| T10 | Data analyst, formal elementary teacher | M.A. | Elementary education, human-computer interaction | 10 |

The research process (**Figure 1**) began with developing interview questions based on two theoretical frameworks: TPACK and TAM. The interview protocol included questions about teachers' understanding of Video GenAI, their perceptions of its educational potential, their ideas for implementing it in teaching practices, and their concerns about potential challenges. Before conducting interviews, we obtained IRB approval and followed appropriate procedures for participant recruitment and consent.

Individual interviews were conducted online via video conferencing software, each lasting approximately one hour. Participants were first provided with the educational guide (See Research Tools section) and access to Video GenAI, allowing them to experiment with video generation capabilities before the interview. During the interviews, which were recorded with participant consent, teachers could reference their hands-on experience with the tool while responding to questions. While following the semi-structured protocol, we allowed for flexible exploration of emerging themes and topics raised by participants.

All interview recordings were transcribed and analyzed through qualitative coding procedures, including initial open coding followed by axial coding, to identify key themes and patterns. This systematic analysis process enabled the researchers to derive meaningful implications regarding the educational potential and implementation considerations of Video GenAI in teaching practices (Details on **Data Collection and Data Analysis Section**).

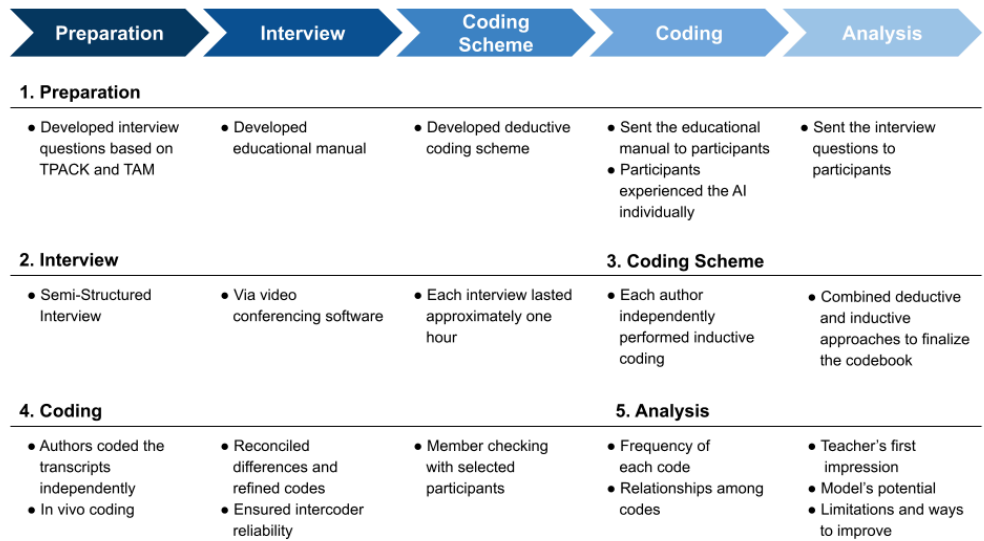

**Figure 1**

*Research Procedure*

**Research Tools**

This study employed three primary research tools: (1) an interview questionnaire based on TPACK and TAM frameworks, (2) Luma Dream Machine (a video generation AI platform), and (3) a comprehensive guide for educational applications of Video GenAI.

The interview questionnaire was developed based on theoretical frameworks examining teachers' TPACK and TAM in relation to AI tools (**Appendix 1**). The questionnaire consisted of three main sections: basic information about participants, main interview questions exploring participants' perceptions and experiences with Video GenAI, and open-ended questions for additional insights.

For the technical tool, we utilized Luma Dream Machine (Luma AI, 2024) (**Figure 1**), which was the most advanced publicly available video generation AI at the time of data collection. While OpenAI's Sora (OpenAI, 2024) was announced during our research period and demonstrated superior capabilities, access was limited to select users. During the interviews, participants were provided access to this tool, allowing them to experiment with video generation capabilities and ground their responses in actual experience rather than theoretical speculation.

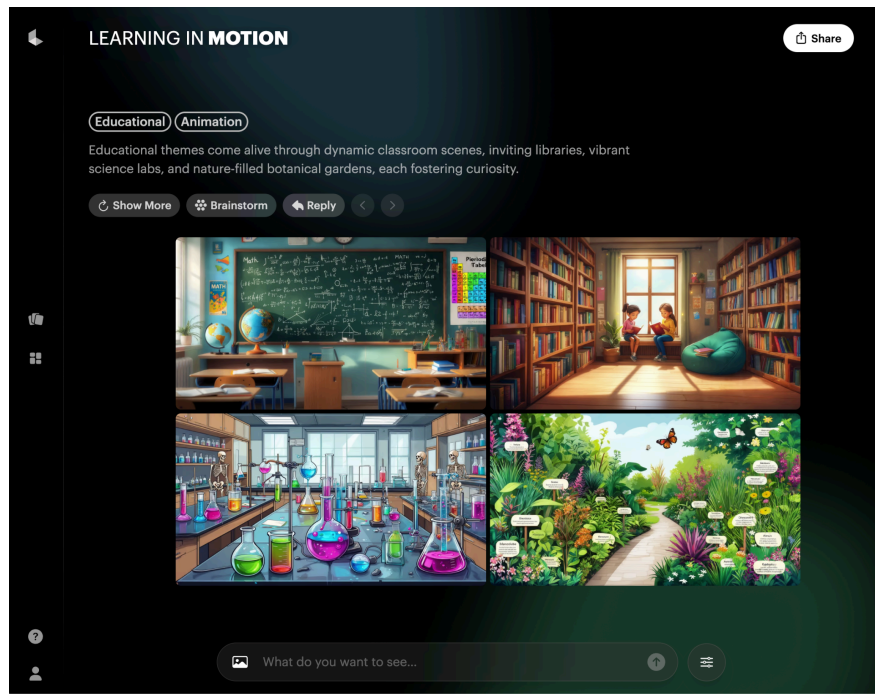

**Figure 2**

*Video Generation Results of Luma Dream Machine*

To support teachers' understanding and use of the video generation AI, we developed a comprehensive educational guide (**Appendix 2**). This guide included detailed instructions for using Luma Dream Machine, from basic login procedures to advanced prompt engineering techniques. More importantly, it provided subject-specific examples of educational applications across various disciplines including English, Science, Art, Social Studies, and Korean Language.

For each subject area, the guide included specific teaching contexts, example prompts, and sample videos, demonstrating how the technology could be integrated into different educational scenarios. The guide also included advanced usage tips for creating more sophisticated outputs, such as detailed character descriptions, setting specifications, and camera angle instructions.

**Data Collection and Data Analysis**

Our data analysis combined both deductive and inductive strategies, reflecting a balance between etic and emic perspectives. Initially, we drew on TPACK and TAM constructs to create a preliminary codebook, defining codes tied directly to theoretical domains such as technological knowledge, pedagogical strategies, content alignment, perceived usefulness, and perceived ease of use. This deductive approach ensured that our analysis was grounded in established theoretical constructs relevant to teachers' integration of new technologies.

Next, the analysis incorporated inductive coding to capture emergent themes that were not fully explained by TPACK and TAM alone. As each researcher independently coded a subset of transcripts, new codes were proposed to account for unique phenomena encountered in participants' responses—such as unexpected ethical considerations, novel curriculum integration ideas, or unique institutional constraints. These new codes were then discussed in weekly team meetings. Discrepancies were resolved through consensus, and the codebook was refined to integrate both deductive and inductive elements, allowing theoretical frameworks to guide interpretation while acknowledging new insights from the data.

We employed a structured coding and thematic analysis approach (Braun & Clarke, 2006; Corbin & Strauss, 1990). The process included open coding to identify initial concepts, axial coding to organize related codes into categories, and thematic coding to develop higher-level themes. Throughout, TPACK and TAM constructs served as interpretive anchors, enabling us to understand participants' responses in relation to known factors affecting technology integration. At the same time, integrating inductive categories allowed the analysis to go beyond these frameworks, unveiling novel perspectives on the educational use of Video GenAI.

Multiple verification strategies enhanced the trustworthiness of the analysis. Five researchers coded the transcripts independently, then met to reconcile differences and refine codes, ensuring intercoder reliability. In-vivo coding was used where possible to honor participants' original language. We also conducted member checking with selected participants to validate our thematic interpretations. Detailed memos documented analytical decisions, ensuring transparency and reproducibility of the coding process.

By integrating TPACK and TAM into both the design and analysis phases while remaining open to emergent themes, we ensured that the resulting interpretations captured both theoretically anticipated aspects and context-specific nuances of teachers' perceptions. This comprehensive and reflective approach enabled a nuanced understanding of educators' engagement with video generation AI, informed by both established theoretical lenses and the realities of participants' experiences.

## Findings

Each teacher independently used Luma Dream Machine, guided by a distributed educational manual that included example prompts. The interview began with general questions about the teachers' overall perceptions of the model, gradually progressing to explore its potential and limitations in the educational settings. To address the research questions within the frameworks of TPACK and TAM, interview data were analyzed in three categories: (1) teachers' impressions and experiences with usage, (2) the potential of the models for educational applications, and (3) future development directions considering current limitations.

A total of three major themes were generated to address the four research questions. Specifically, the final coding structure consists of 1 major theme, 3 sub-themes, and 6 codes under RQ1, 1 major theme, 4 sub-themes, and 22 codes under RQ2, and 1 major theme, 2 sub-themes, and 7 codes under RQ3 (Details are in the **Appendix 3**).

## RQ1. What are teachers' first impressions and usage experiences of Video GenAIs?

The adoption of AI in education depends on how teachers perceive and engage with it. Their initial impressions shape expectations, while hands-on experience with usability and usefulness refines their views. These factors ultimately influence their willingness to integrate AI-driven tools into classrooms. Understanding this process helps identify both opportunities and challenges in implementing Video GenAI.

### *Impression*

Teachers' impressions were categorized into positive and negative groups, with the frequency of positive impressions being approximately three times higher than that of negative ones. Teachers with a positive outlook on Video GenAI perceived it as a remarkable innovation —akin to watching a video recorded with a real camera—and exhibited high levels of expectancy for its application in educational settings. They expressed satisfaction with the models' attributes, such as rapid generation times and the realistic quality of the content, and were optimistic about leveraging the technology to enhance their teaching capabilities.

Conversely, teachers who held negative impressions highlighted their dissatisfaction with the generated content, raising concerns about the technology's uncertainty and potential limitations in educational contexts. These educators voiced apprehensions regarding possible side effects and misuse of the technology. *"The generated video was like a movie filmed with an expensive camera; however, it created negative scenes as realistically as others."* Although both groups interacted with the same model, their evaluations of content quality differed significantly, likely reflecting the variations in their expectations, experiences, and technological readiness.

Teachers' impressions of the model reflect their overall judgments or perceived value of the product. While these impressions were broadly categorized into positive and negative factors, individual teachers appeared to shape their overall judgments by considering the balance between perceived gains and losses (Zhang et al., 2022). These personal judgments, grounded in their understanding of the model, contribute to their TK, which subsequently influences the development of their TPACK (Koehler & Mishra, 2009). Therefore, examining the factors shaping their perceived gains and losses, derived from their utilization experiences, is essential to understanding technological integration in education.

### *Utilization Experience*

Teachers' individual experiences with using the Video GenAI were categorized into three areas: perceived ease of use, perceived usefulness, and prior experience with similar technologies. Just as their impressions of the model spanned a spectrum of positive and negative evaluations, their assessments of ease of use and perceived usefulness also reflected both strengths and limitations (Davis, 1989).

Regarding perceived ease of use, they highlighted features such as the intuitive interfaces and the model's ability to analyze natural language and image sources as notable advantages. However, they also identified key shortcomings, including difficulties in processing the Korean language, challenges in generating content that aligned precisely with user intentions, and the lack of clear prompt guidelines for novice users. Meanwhile, teachers perceived the generation time differently: some thought it was fast enough, while others considered it too slow for classroom use. This ambivalent perception may stem from subjective variations in speed perception among individuals, influenced by comparisons to prior experiences with similar technologies, individual differences in cognitive processing speed, and familiarity with digital tools (Card et al., 2018; Tversky & Kahneman, 1974).

For perceived usefulness, teachers placed high value on the model's potential applications in classroom settings, including facilitating subject-specific activities, promoting AI literacy, supporting students' creative activities, and reducing preparation time for instructional materials. *"No matter how much they make public service advertisements, they don't achieve visible results. When these are produced at the desired level, it'll be very useful for sharing with parents."* Conversely, they raised concerns about the restricted capacity of free accounts and stressed the critical role of teacher oversight in ensuring safe and effective use by students. *"It may not be easy for teachers to review if the wrong video is generated or if it diverges from the intended direction."* while Video GenAI show significant promise as educational tools when applied thoughtfully and within a well-structured framework, teachers expressed reservations about their immediate adoption due to unresolved challenges and uncertainties.

Since Video GenAI is a new technology with limited information on educational implementation, teachers relied on their prior knowledge of AI platforms to envision potential applications in classroom settings. Although their evaluations were based on hypothetical scenarios—and therefore subject to some degree of inaccuracy—this exercise highlighted the potential reactions or resistance from educators that the tool might encounter before its adoption in classroom environments. Individual experiences with AI tools significantly influenced the evaluation of each model, ultimately shaping varying levels of willingness to adopt and rates of utilization.

Although the three categories (perceived ease of use, perceived usefulness, and prior experience with similar technologies) were distinct, they were closely related. For instance, the evaluation of usefulness was often influenced by teachers' perceptions of ease of use (Davis, 1989). Similarly, their assessments of both ease of use and perceived usefulness were frequently shaped by comparisons with their prior experiences using artificial intelligence platforms (Venkatesh & Davis, 2000; Venkatesh et al., 2003). The interplay among these factors underscores the significance of ease of use and perceived usefulness in evaluating the model's educational applicability, with specific criteria rooted in prior experiences with similar AI tools.

### Willingness to use

Notably, most of the interviewees were educators at the forefront of integrating AI into education, with extensive experience in using diverse educational tools. This background contributed to a high level of willingness to adopt the model. The proportion of teachers expressing positive intentions to use the technology was 2.3 times greater than those indicating reluctance. Many participants identified its potential as a motivational tool for engaging students or as a creative source in project-based learning. *"It may vary slightly depending on the topic, but if there is a project-based class, I think I will use it once at the end."* Overall, these teachers demonstrated strong agency in adopting new technologies, attributed to their experience in integrating various resources and tools to meet educational needs.

However, a significant number of teachers also expressed concerns about the model, reflecting a low willingness to adopt it. *"I am somewhat skeptical about whether this technology should be broadly introduced in primary schools."* Specific issues included the model's tendency toward body distortions in generated outputs, which were perceived as potentially harmful to young students' perceptions.

The difference in teachers' behavioral intentions can be attributed to their varying impressions and perceptions of the model. According to the TAM model, perceived ease of use and perceived usefulness shape attitudes toward using a system, which ultimately influence behavioral intentions (Davis, 1989). While individual differences will persist, enhancing perceived ease of use and usefulness—key determinants of willingness to adopt technology (Davis, 1989; Venkatesh & Davis, 2000)—is essential for encouraging educational applications. Moreover, the systems' technical stability plays a significant role in shaping users' functional and emotional experiences (Zhang et al., 2022). Addressing both technical and pedagogical challenges will be crucial to fostering teachers' agency and willingness to adopt the model effectively.

## RQ2. What potential do Video GenAIs have for educational applications?

Video GenAI emerged as a promising technology with the capacity to reshape classroom instruction through multimedia content generation. To explore how teachers perceive its educational potential, we investigated their experiences and observed practical examples of Video GenAI integration in K-12 settings. Interviews with the leading teachers revealed their perspectives on Video GenAI's role in enhancing subject comprehension, fostering creativity, and diversifying instructional strategies.

### Enhancing comprehension of the course

Teachers highlighted how Video GenAI can enhance course comprehension, emphasizing two types of application: (1) assisting teachers in creating supplementary materials, and (2) serving as a learning tool for students to solidify their understanding.

In the first application, teachers noted that Video GenAI can generate diverse, contextually accurate, and dynamic visuals with relatively little time and effort. This feature allows them to create supplementary materials that effectively visualize abstract or complex scenarios, making them more accessible to students. As they pointed out, explaining certain concepts often requires explicit elaboration that can be challenging through conventional means. One teacher said, *"It could be used in science classes to visualize experiments and things like that."* Another teacher mentioned, *"In history classes, if there are records that are now only in text or are hard to find on YouTube... I thought it would be nice to make a specific request for this and make some reenactment videos that are not on YouTube."* Such views are consistent with prior examples of using AI for automating PPT-to-video conversions (Chen et al., 2023) and integrating AI-generated images with scripts (Weerakoon et al., 2024). However, participants recognized that current Video GenAI tools were limited in visual naturalness and diversity, requiring significant editing to refine and generate contextually accurate and dynamic content.

Teachers also observed that Video GenAI may enable personalized explanations, allowing for easy customization of learning materials. As one stated, *"I believe we may create a material that assists students in comprehending subjects through a personalized way."* They envisioned tailoring video content to students with various levels of prior knowledge or learning styles. In addition, some teachers mentioned that turning literary or non-literary texts into short videos could improve engagement in language arts classes. One participant described, *"If you are teaching in Korean class a literary or non-literary work and you lack a suitable video to inspire students, but only have some decent images, you might create a little inspiring videos from those images."*

In the second application, Video GenAI can be used by students themselves to reinforce subject knowledge. Through generating videos that illustrate what they have learned, students may deepen their understanding of complex material. This could be particularly beneficial in subjects requiring strong visualization, such as science or history. They explained that, *"While there are many resources like Google Earth and historical photographs, students often lack access to dynamic, immersive video content that enables them to bring past events to life."* Another teacher noted that the technology could help compensate for limitations in laboratory facilities by showing experimental outcomes or simulating phenomena. These visual representations and spatial reasoning play a crucial role in facilitating the understanding of scientific concepts and enhancing learning (Evagorou et al., 2015; Ramadas, 2009; Tippett, 2016). Across these scenarios, participants reported that creating and watching offer students with more concrete ways of internalizing learning content.

### *Promotion of Creative Expression and Divergent Thinking*

Teachers commonly reported that Video GenAI supports students' creative expression and divergent thinking. They identified three broad functions: (1) supporting imaginative expression, (2) encouraging multiple perspectives or outcomes during creative activities, and (3) enhancing composition skills.

First, participants noted that Video GenAI enables students to portray imaginative scenes—whether realistic or fantastical—that are difficult to film. One teacher stated, *"Although students can already express themselves through writing or speaking, from an elementary school student's perspective, writing can feel like a highly complex symbolic system. In that sense, the ability to instantly produce visual materials can help expand their imagination."* Another teacher added, *"Children often imagine things like meeting a dragon or interacting with it. If their imagined scenarios could be turned into a video and shared, that opens up a new kind of usefulness—something that previously didn't exist."*

Second, they highlighted how rapidly generated video outputs let students explore various possibilities in a short time, spurring broader thinking. *"Instead of just providing the previous scene as text when kids write a novel, what if we present it as a video? This could stimulate their imagination and encourage them to think about the next part of the story, helping them expand their thinking in a more creative and divergent way."* By experimenting with a variety of prompts and styles, learners might expand their creative scope beyond a single, fixed interpretation.

Third, several teachers described how translating ideas into specific prompts may refine students' language usage. *"When students concretize their imagined stories or images through sentences ... I believe this helps improve their ability to formulate sentences."* Because GenAI requires precise text descriptions, students might learn to construct more detailed statements and refine their linguistic expression  (Lee et al., 2023). Participants felt that sharing prompts and outputs between peers further enhances these gains, as learners can quickly see how differences in wording lead to different video results.

### *Innovation in Teaching and Learning Methods*

Teachers also emphasized how Video GenAI could transform various teaching and assessment methods. Within this theme, the role of Video GenAI in supporting project-based learning (PBL), diversifying assessment formats, and its relationship with learners' AI literacy were mainly discussed.

First, participants envisioned using Video GenAI to enrich PBL activities by providing visual representations in students' projects. One teacher remarked, *"Project-based learning inherently includes self-directed learning. Students can use AI-generated videos to convey their knowledge and concepts to others. This approach seems highly beneficial for deepening understanding and engagement."* In PBL, students typically create presentations or develop prototypes to communicate their ideas and findings (Notari et al., 2014; Owen & Hite, 2022; Shashidhara et al., 2024). Teachers felt Video GenAI could further enhance storytelling or demonstrate complex ideas. In some cases, teachers said it could even become the centerpiece of a project, with students creating entirely novel stories or short films. *"At first, the students were simply curious about how AI-generated videos worked. But as they experimented with different tools, they started thinking critically about composition, lighting, and transitions—just like real filmmakers."*

Second, teachers described ways in which video outputs might be used for formative and summative evaluations. They noted that providing AI-generated prompts or final videos could serve as an additional data source to assess student learning. *"If students submit their AI-generated prompts, we could use them for diagnostic, formative, and summative assessments, making the evaluation process more comprehensive."* By including creative projects alongside traditional tests, teachers felt they could better gauge students' progress and problem-solving skills, like a prior study (Lee et al., 2023) that used similar approaches with image generation tools. In their view, these methods would reduce reliance on standardized exams and encourage students to demonstrate learning in more innovative ways.

Third, a number of teachers underscored the relevance of AI literacy when implementing Video GenAI in educational settings. One stressed, *"Before using generative AI, students should first receive lessons on AI ethics and responsible usage."* Teachers suggested that instruction on how the AI processes data, potential biases, and privacy issues must precede hands-on activities, while some proposed integrating that content into existing subjects rather than allocating separate class hours. This indicates varying views on the best approach, but general consensus on the need to ensure students understand the ethical and practical dimensions of Video GenAI (Annapureddy et al., 2024; Sullivan et al., 2024). On the other hand, some participants anticipated that through classes utilizing Video GenAI, learners would "gain experience interacting with AI and gain insight," ultimately improving their AI literacy.

**RQ3. What are the limitations of Video GenAIs in educational applications, and what are the ways to improve them and future development directions?**

The limitations of Video GenAI in education can be grouped into technical issues and policy challenges. Technical issues include performance limitations, language barriers, and safety concerns. Policy challenges involve age restrictions, copyright issues, and unclear guidelines.

*Technical issues and limitations*

The Video GenAI models face several technical challenges that undermine their effectiveness in educational applications. These models, while undergoing significant technological advancements, have developed more slowly compared to text- or image-based AI technologies (Bhagwatkar et al., 2020; Liu et al., 2024a). The complexity of spatio-temporal modeling and the scarcity of videotext datasets are primary factors impeding their development (Feurriegel, 2024; Guo et al., 2024). These technical issues and challenges lead to difficulties and tangible implications for the use of these models in educational activities within the classroom context, where various complexities must be carefully considered.

First, performance issues are a notable limitation. The teachers frequently cited prolonged rendering times and the short length of generated videos as significant challenges. The constraints make it challenging to scale these tools for widespread classroom adoption. Similar concerns have been raised in the broader field of GenAI (Mittal et al., 2024). Teachers also expressed concerns about the quality of generated videos. Inconsistencies across frames, such as disjointed scene transitions and unnatural human representations—particularly in hands and facial expressions—undermine the utility of these tools for creating cohesive educational content. Similar issues have also been observed in the use of image-based GenAI for art education (Lee et al., 2023).

Additionally, teachers have raised concerns regarding issues stemming from biased data or the potential for hallucinations, as these models can reinforce students' misconceptions or biases, and even lead to discrimination. These problems may be particularly pronounced in subjects such as science and history, where accurate delivery of content is crucial (Baker & Hawan, 2022). As one educator noted, *"If I try to generate a video explaining the orbit of our solar system, and even if I ask it to be based on facts, it could still produce unhelpful content."* Furthermore, unintended or irrelevant scenes often generated by the model hinder the overall quality and applicability of the output, further limiting its effectiveness in educational contexts.

Second, limited language support has been identified as another barrier, particularly with the model's restricted ability to handle non-English prompts, especially Korean. This shortcoming reflects broader limitations in supporting diverse languages, which is critical for leveraging these tools in global educational contexts (Vayani et al., 2024). This limitation could also lead to a decline in learning motivation, as non-English-speaking students may have to spend excessive time translating their prompts into a foreign language.

Third, safety is another major concern. The absence of robust filtering mechanisms allows for the potential generation of harmful or inappropriate content (Wang et al., 2024), which in turn raised concerns about the misuse of these tools among teachers. As one educator noted, *"There is a high possibility that, in the end, they will use both video and audio to generate deep fake content using influencers' images."* Experts highlighted the need for built-in safety filters to mitigate these risks and maintain a secure learning environment. One expert acknowledged the challenge stating, *"Defining the criteria for abusive or suggestive content may be difficult; however, blocking such images in a general sense would be sufficient."*

Lastly, the lack of effective communication features within these tools was identified as a factor that further restricts the practical use of video generation models in education. Current Video GenAI tools like Luma Dream Machine (Luma AI, 2024) are primarily designed for professional content creators rather than educational purposes. Teachers noted the difficulty of distributing generated content and prompts without relying on external platforms. This fragmentation complicates workflows and reduces efficiency. Additionally, the absence of teacher monitoring functions was seen as a critical gap. Without the systems that link teacher and student accounts or enable real-time oversight, it becomes challenging to ensure that student-generated content aligns with educational goals (Aly, 2024; Marouf et al., 2025). One teacher said that *"I worry that without proper oversight, students might misuse the tool or create content that deviates from the learning objectives. We need a system where teachers can track progress and provide guidance at key stages."*

### Policy Issues and Limitations

The integration of Video GenAI in K-12 education faces significant policy and institutional challenges. One pressing issue is the stringent age restrictions associated with GenAI tools, which limit their applicability across various school levels (Shepperd, 2024; UNESCO, 2023). Some educators suggested that *"age restrictions should be relaxed so that even fourth graders can utilize the technology effectively."* These administrative barriers highlight the need for policy adjustments to ensure equitable access across all grades.

Parental consent requirements and legal implications further complicate the adoption process (UNESCO, 2023). Educators reported that *"obtaining parental consent is time-consuming and administratively burdensome,"* especially in contexts where alternative instruction must be provided for non-consenting students. Even when consent is obtained, *"there remains the risk of future complaints or lawsuits"* creating hesitation among teachers to fully integrate these tools into their curriculum.

Another significant concern lies in the lack of robust legal safeguards. Teachers emphasized the importance of addressing intellectual property rights and copyright issues, noting that *"AI-generated outputs might not entirely belong to the creator, which complicates ownership and usage rights."* The absence of clear legal guidelines could lead to negligence and infringement of personal rights (Shen, 2024). With the rapid development of various GenAI programs, students can now create art pieces with minimal effort, which may lead them to undervalue others' creative work. Moreover, the potential misuse of GenAI tools poses ethical challenges, particularly in ensuring the safety and appropriateness of generated content for educational purposes.

The absence of clear and standardized guidelines further undermines confidence in the use of Video GenAI. While some fragmentary guidelines exist (KERIS, 2022), they are often limited in scope and lack enforceable authority. Teachers pointed out that *"national-level guidelines must be developed to provide clear direction on ethical and pedagogical usage."* This would help alleviate the current reliance on individual judgment, which can lead to inconsistent and sometimes ineffective application of the technology.

Infrastructure deficiencies also present substantial barriers (Zhao & Frank, 2003). Teachers highlighted the inadequacy of school resources, noting that *"many schools lack*

*sufficient devices and reliable internet access to support AI tools."* These challenges are further compounded by the lack of technical support staff, which makes managing and maintaining the necessary infrastructure difficult. As a result, the learning gap in technology exposure will continue to grow depending on the educational environment (Van Dikk, 2005). One expert also warned, *"If the private education sector progressively implements new technologies while public schools do not, the gap between public and private education will only intensify."* Addressing these gaps in infrastructure will be essential to ensure the widespread adoption of Video GenAI in classrooms.

Financial and administrative hurdles further hinder adoption. The high cost of AI tools, including subscription fees, presents a significant barrier for schools with limited budgets. Teachers stated that *"government funding or subsidies for software costs would greatly facilitate adoption."* Additionally, administrative challenges, such as overseas payment restrictions, make it difficult for schools to access advanced AI tools, highlighting the need for streamlined administrative processes.

To overcome these challenges, educators emphasized the importance of professional development and training programs to build teachers' confidence and technological literacy. They also stressed the need for student and parental education on ethical considerations, such as *"understanding copyright issues and ensuring responsible use of the tools."* Comprehensive policy frameworks, improved infrastructure, and targeted financial support will be critical to realizing the full potential of Video GenAI in K-12 education while ensuring ethical and equitable practices (Ertmer, 1999).

# Discussion

## *Implications*

The findings of this study underscore the transformative potential of Video GenAI in K-12 education. Teachers identified substantial opportunities for enhancing classroom instruction, fostering creativity, and facilitating authentic task design. Specifically, Video GenAI can improve student comprehension through subject-specific visualizations, enable personalized learning experiences, and support creative expression. These applications align with contemporary educational priorities, including differentiated instruction, project-based learning, and the integration of technology to address diverse learner needs.

Notably, these findings resonate with Chen et al. (2024) and Lee et al. (2023), who demonstrated how GenAI supports creative expression in K-12 context. However, unlike prior text-centric AI studies (e.g., Lo, 2023), our research highlights the added visual dimension of Video GenAI, reinforcing the importance of TPK when designing multimodal tasks. By allowing teachers and students to quickly visualize otherwise abstract concepts, Video GenAI offers a novel mechanism for improving clarity, engagement, and innovation in daily classroom activities.

From a pedagogical standpoint, the study also highlights the need for structured approaches to implementing Video GenAI in classrooms. Teachers emphasized the importance of explicit learning objectives, guided practice, and scaffolding to ensure meaningful integration. Moreover, the technology's potential to reduce teacher preparation time could be particularly valuable in resource-constrained educational settings. This aligns with previous studies on text-based GenAI (Adeshola & Adepoju, 2024; Memarian & Doleck, 2023; Opara et al., 2023), which have also emphasized the importance of structured implementation for effective integration. This suggests that while video and text are different modalities, Video GenAI, like text-based GenAI, shares foundational considerations for educational use beyond modality differences. Therefore, a unified approach to researching their pedagogical implications is necessary.

The study also revealed significant barriers to adoption, most notably technical challenges, ethical concerns, and a lack of institutional support. The frequent inconsistencies and hallucinations in generated outputs, limited support for non-English languages, and absence of built-in safety mechanisms are technical hurdles that must be addressed. Previous research which used image GenAI (Chen et al,. 2024; Lee et al., 2023) also reported that image GenAI systems face similar issues, further hindering their adoption. To address these challenges, ethical concerns can be mitigated by implementing safety guardrails and integrating toxic content detection models (Chi et al., 2024; He et al., 2024; Ousidhoum et al., 2021). For inconsistencies and hallucinations, recent advancements like ControlNet-based consistency techniques (Zhang et al., 2023; Zhao et al., 2024), guidance (Ho & Salimans, 2021; Bansal et al., 2023) and Retrieval Augmented Generation (RAG) (Arefeen et al., 2024; Lewis et al., 2020) should be leveraged. Most importantly, multilingual support necessitates the development of Video GenAI trained on diverse languages.

***Policy and Ethical Considerations***

Given the potential for inappropriate or biased content generation, policymakers and school administrators must collaborate with developers to implement robust content filters and age-based usage guidelines. Age restrictions in GenAI tools, in particular, pose obstacles for younger learners—yet loosening these restrictions without proper safety measures could expose students to harmful materials (Holmes & Miao, 2023; Lim et al., 2023). Clear regulatory frameworks and enforceable national-level policies (e.g., UNESCO, 2023) are therefore essential to standardize ethical protocols around data privacy, intellectual property, and the responsible deployment of AI in schools.

Teacher training should also include explicit instruction on how to critically evaluate and guide students in the responsible use of Video GenAI. Without sufficient AI literacy, students may either overtrust AI outputs or be unaware of potential biases embedded within generative models (Ausat et al., 2023; Trust et al., 2023). Finally, to uphold student safety and academic integrity, institutions must establish transparent rules regarding the storage and use of student-generated data, particularly when combining AI outputs with personal information or other classroom resources.

***Limitations and Future Directions***

While this study provides valuable insights into the potential and challenges of Video GenAI, it does have limitations that suggest avenues for further research. First, our focus on "leading teachers", professionals already well-versed in educational technology, may mean their perceptions are more positive than those of the general teaching population. Including a broader demographic sample of educators could yield a more representative view of barriers to adoption.

Second, the technology itself is rapidly evolving. Several technical limitations identified here (e.g., rendering speed, hallucinations, unstable outputs) could quickly become obsolete as Video GenAI advances. Future studies should adopt longitudinal or iterative designs to capture how improvements in model performance might shift teacher perceptions, usability, and ethical considerations over time.

Third, further research is needed to assess the sustained impact of AI-driven video content on student learning, creativity, and critical thinking. Investigation into how Video GenAI can address diverse cultural and linguistic contexts is especially critical, given the challenges identified with non-English languages. Comparative cross-regional or cross-grade studies would refine our understanding of how to tailor best practices for integrating Video GenAI into various classroom settings.

Lastly, professional development programs that build teachers' TPACK and TAM specifically for Video GenAI remain an essential area of inquiry. Designing empirically tested teacher-training curricula, complete with hands-on practice, prompt engineering workshops, and ethical-usage guidelines, will be pivotal for maximizing the educational benefits of this technology.

## Conclusive Remark

This study provides a foundational understanding of the educational applications of Video GenAI from the perspectives of leading K-12 teachers, using the TPACK and TAM frameworks as analytical lenses.  The findings demonstrate that, while Video GenAI holds the potential to revolutionize teaching and learning, its successful integration requires overcoming significant technical, ethical, and institutional hurdles. By illuminating teachers' perceptions and needs, this research contributes to the broader discourse on AI-driven education and offers practical implications for policy, teacher training, and technology development. From a TPACK perspective, effective pedagogical strategies must be identified to facilitate the meaningful integration of Video GenAI despite existing constraints. This study underscores the need for further research and discussion on instructional design principles and models that can guide the integration of Video GenAI while accounting for its dynamic nature and the diverse contexts of educational settings.

As Video GenAI continues to evolve, its capacity to enhance educational equity, creativity, and engagement remains vast. However, realizing these benefits depends on addressing the barriers identified in this study, ranging from technical inconsistencies to ethical concerns, and building a supportive ecosystem that empowers teachers while protecting learners. More importantly, this study highlights the need to move beyond merely adopting new technology and instead focus on identifying optimal instructional strategies and design principles that integrate Video GenAI as a meaningful pedagogical tool. By actively interpreting technology as an integrated component of educational goals, methods, and instructional design knowledge—rather than as an external add-on—teachers can harness its full potential. Through collaboration among stakeholders and a commitment to responsible innovation, the educational community can leverage Video GenAI to create transformative, inclusive, and pedagogically sound learning experiences for all students.

**References**

Adeshola, I., & Adepoju, A. P. (2024). The opportunities and challenges of ChatGPT in education. *Interactive Learning Environments, 32*(10), 6159-6172. http://dx.doi.org/10.1080/10494820.2023.2253858

Adetayo, A. J., Enamudu, A. I., Lawal, F. M., & Odunewu, A. O. (2024). From text to video with AI: the rise and potential of Sora in education and libraries. *Library Hi Tech News*. http://dx.doi.org/10.1108/LHTN-02-2024-0028

Ahmad, Z., Kaiser, W., & Rahim, S. (2023). Hallucinations in ChatGPT: An unreliable tool for learning. *Rupkatha Journal on Interdisciplinary Studies in Humanities, 15*(4), 12-29. https://doi.org/10.21659/rupkatha.v15n4.17

Alasadi, E. A., & Baiz, C. R. (2023). Generative AI in education and research: Opportunities, concerns, and solutions. *Journal of Chemical Education, 100*(8), 2965-2971. http://dx.doi.org/10.1021/acs.jchemed.3c00323

Almuhanna, M. A. (2024). Teachers' perspectives of integrating AI-powered technologies in K-12 education for creating customized learning materials and resources. *Education and Information Technologies*, 1-29. http://dx.doi.org/10.1007/s10639-024-13257-y

Alwaqdani, M. (2024). Investigating teachers' perceptions of artificial intelligence tools in education: potential and difficulties. *Education and Information Technologies*, 1-19. http://dx.doi.org/10.1007/s10639-024-12903-9

Aly, M. (2024). Revolutionizing online education: Advanced facial expression recognition for real-time student progress tracking via deep learning model. *Multimedia Tools and Applications*, 1-40. http://dx.doi.org/10.1007/s11042-024-19392-5

Annapureddy, R., Fornaroli, A., & Gatica-Perez, D. (2024). Generative AI literacy: Twelve

 defining competencies. *Digital Government: Research and Practice*.

 http://dx.doi.org/10.1145/3685680

Arefeen, M. A., Debnath, B., Uddin, M. Y. S., & Chakradhar, S. (2024). ViTA: An Efficient

 Video-to-Text Algorithm using VLM for RAG-based Video Analysis System. In

 *Proceedings of the IEEE/CVF Conference on Computer Vision and Pattern Recognition*

 (pp. 2266-2274). http://dx.doi.org/10.1109/CVPRW63382.2024.00232

Ausat, A. M. A., Massang, B., Efendi, M., Nofirman, N., & Riady, Y. (2023). Can chat GPT

 replace the role of the teacher in the classroom: A fundamental analysis. *Journal on

 Education, 5*(4), 16100-16106. https://doi.org/10.31004/joe.v5i4.2745

Ayeni, T. P. (2024). Educational and Professional Development Approaches in a Digital World:

 Lesson from the COVID-19 Pandemic in Africa. *Beyond the Chalkboard: Crafting

 Strategies for Human Capital Development in the Digital World*, 281-307.

Baker, R. S., & Hawan, A. (2022). Algorithmic bias in education. *International Journal of

 Artificial Intelligence in Education, 32*, 1052-1092.

 https://doi.org/10.1007/s40593-021-00285-9

Bansal, A., Chu, H. M., Schwarzschild, A., Sengupta, S., Goldblum, M., Geiping, J., &

 Goldstein, T. (2023). Universal guidance for diffusion models. In *Proceedings of the

 IEEE/CVF Conference on Computer Vision and Pattern Recognition* (pp. 843-852).

 http://dx.doi.org/10.1109/CVPRW59228.2023.00091

Bar-Tal, O., Chefer, H., Tov, O., Herrmann, C., Paiss, R., Zada, S., Ephrat, A., Hur, J., Liu, G.,

 Raj, A., Li, Y., Rubinstein, M., Michaeli, T., Wang, O., Sun, D., Dekel, T., & Mosseri, I.

 (2024, December). Lumiere: A space-time diffusion model for video generation. In *SIGGRAPH Asia 2024 Conference Papers* (pp. 1-11).

http://dx.doi.org/10.1145/3680528.3687614

Berbar, S. (2023, November). User Experience and Technology Adoption: The Mediating Effect of Perceived Ease of Use on Senso-Aesthetic Openness and Behaviour Intention. In *2023 17th International Conference on Signal-Image Technology & Internet-Based Systems (SITIS)* (pp. 38-45). IEEE. http://dx.doi.org/10.1109/SITIS61268.2023.00016

Bhagwatkar, R., Bachu, S., Fitter, K., Kulkarni, A., & Chiddarwar, S. (2020, December). A review of video generation approaches. In *2020 international conference on power, instrumentation, control and computing (PICC)* (pp. 1-5). IEEE. http://dx.doi.org/10.1109/PICC51425.2020.9362485

Bobek, E., & Tversky, B. (2016). Creating visual explanations improves learning. *Cognitive research: principles and implications, 1*, 1-14. http://dx.doi.org/10.1186/s41235-016-0031-6

Brame, C. J. (2017). Effective educational videos: Principles and guidelines for maximizing student learning from video content. *CBE—Life Sciences Education, 15*(4). http://dx.doi.org/10.1187/cbe.16-03-0125

Braun, V. & Clarke, V. (2006). Using thematic analysis in psychology. *Qualitative Research in Psychology, 3*(2), 77-101. http://dx.doi.org/10.1191/1478088706qp063oa

Brown, C., White, R., & Kelly, A. (2023). Teachers as educational change agents: What do we currently know? Findings from a systematic review. *Emerald Open Research, 1*(3). http://dx.doi.org/10.1108/EOR-03-2023-0012

Card, S. K., Moran, T. P., & Newell, A. (2018). *The psychology of human-computer interaction*. Crc-Press. https://doi.org/10.1201/9780203736166

Chau, P. Y. (1996). An empirical assessment of a modified technology acceptance model. *Journal of management information systems, 13*(2), 185-204. http://dx.doi.org/10.1080/07421222.1996.11518128

Chen, S., Liu, Q., & He, B. (2023, November). A Generative AI-based Teaching Material System Using a Human-In-The-Loop Model. In *2023 International Conference on Intelligent Education and Intelligent Research (IEIR)* (pp. 1-8). IEEE. http://dx.doi.org/10.1109/IEIR59294.2023.10391244

Chen, Y., Zhang, X., & Hu, L. (2024). A progressive prompt-based image-generative AI approach to promoting students' achievement and perceptions in learning ancient Chinese poetry. *Educational Technology & Society, 27*(2), 284-305. https://doi.org/10.30191/ETS.202404_27(2).TP01

Chen, Y. C., Lu, Y. L., & Lien, C. J. (2021). Learning environments with different levels of technological engagement: a comparison of game-based, video-based, and traditional instruction on students' learning. *Interactive Learning Environments, 29*(8), 1363-1379. http://dx.doi.org/10.1080/10494820.2019.1628781

Chi, J., Karn, U., Zhan, H., Smith, E., Rando, J., Zhang, Y., Plawiak, K., Coudert, Z. D., Upasani, K., & Pasupuleti, M. (2024). Llama guard 3 vision: Safeguarding human-ai image understanding conversations. *arXiv preprint arXiv:2411.10414*. https://doi.org/10.48550/arXiv.2411.10414

Cong-Lem, N. (2024). Teacher agency for change and development in higher education: A scoping literature review. *International Journal of Educational Reform*, https://doi.org/10.1177/10567879231224744 .

Corbin, J. M., & Strauss, A. (1990). Grounded theory research: Procedures, canons, and

   evaluative criteria. *Qualitative sociology, 13*(1), 3-21.

   http://dx.doi.org/10.1007/BF00988593

Cukurova, M., Miao, X., & Brooker, R. (2023, June). Adoption of artificial intelligence in

   schools: unveiling factors influencing teachers' engagement. In *International conference*

   *on artificial intelligence in education* (pp. 151-163). Cham: Springer Nature Switzerland.

   http://dx.doi.org/10.1007/978-3-031-36272-9_13

Dao, X. Q., Le, N. B., & Nguyen, T. M. T. (2021, March). Ai-powered moocs: Video lecture

   generation. In *Proceedings of the 2021 3rd International Conference on Image, Video and*

   *Signal Processing* (pp. 95-102). http://dx.doi.org/10.1145/3459212.3459227

Davis, F. D. (1989). Perceived usefulness, perceived ease of use, and user acceptance of

   information technology. *MIS Quarterly, 13*(3), 319-340. https://doi.org/10.2307/249008

Davis, F. D., Bagozzi, R. P., & Warshaw, P. R. (1989). User Acceptance of Computer

   Technology: A Comparison of Two Theoretical Models. *Management Sciencei, 35*(8),

   982-1003. https://doi.org/10.1287/mnsc.35.8.982

Deng, L., Turner, D. E., Gehling, R., & Prince, B. (2010). User experience, satisfaction, and

   continual usage intention of IT. *European Journal of Information Systems, 19*(1), 60-75.

   http://dx.doi.org/10.1057/ejis.2009.50

Ertmer, P. A. (1999). Addressing first-and second-order barriers to change: Strategies for

   technology integration. *Educational Technology Research and Development, 47*, 47-61.

   https://doi.org/10.1007/BF02299597

Evagorou, M., Erduran, S., & Mäntylä, T. (2015). The role of visual representations in scientific

   practices: from conceptual understanding and knowledge generation to 'seeing' how science works. *International journal of Stem education, 2*, 1-13.

http://dx.doi.org/10.1186/s40594-015-0024-x

Feuerriegel, S., Hartmann, J., Janiesch, C., & Zschech, P. (2024). Generative ai. *Business & Information Systems Engineering, 66*(1), 111-126.

https://doi.org/10.1007/s12599-023-00834-7

Galindo-Domínguez, H., Delgado, N., Campo, L., & Losada, D. (2024). Relationship between teachers' digital competence and attitudes towards artificial intelligence in education. *International Journal of Educational Research, 126*, 102381.

http://dx.doi.org/10.1016/j.ijer.2024.102381

Gao, S., Zhou, P., Cheng, M. M., & Yan, S. (2023). Masked diffusion transformer is a strong image synthesizer. In *Proceedings of the IEEE/CVF International Conference on Computer Vision* (pp. 23164-23173). http://dx.doi.org/10.1109/ICCV51070.2023.02117

Gaudin, C., & Chaliès, S. (2015). Video viewing in teacher education and professional development: A literature review. *Educational research review, 16*, 41-67.

http://dx.doi.org/10.1016/j.edurev.2015.06.001

Giannakos, M. N., Jaccheri, L., & Krogstie, J. (2015). How video usage styles affect student engagement? implications for video-based learning environments. In *State-of-the-Art and Future Directions of Smart Learning* (pp. 157-163). Singapore: Springer Singapore.

http://dx.doi.org/10.1007/978-981-287-868-7_18

Google DeepMind. (2024). *Veo 2*. https://deepmind.google/technologies/veo/veo-2/

Granić, A. (2022). Educational technology adoption: A systematic review. *Education and Information Technologies, 27*(7), 9725-9744.

http://dx.doi.org/10.1007/s10639-022-10951-7

Guo, X., Liu, J., Cui, M., & Huang, D. (2024). I4VGen: Image as Stepping Stone for Text-to-Video Generation. *arXiv preprint arXiv:2406.02230.*

Gutiérrez-González, R., Zamarron, A., & Royuela, A. (2024). Video-based lecture engagement in a flipped classroom environment. *BMC Medical Education, 24*(1), 1218. http://dx.doi.org/10.1186/s12909-024-06228-x

Halaweh, M. (2023). ChatGPT in education: Strategies for responsible implementation. *Contemporary educational technology, 15*(2), 421-431. http://dx.doi.org/10.30935/cedtech/13036

He, X., Zannettou, S., Shen, Y., & Zhang, Y. (2024, May). You only prompt once: On the capabilities of prompt learning on large language models to tackle toxic content. In *2024 IEEE Symposium on Security and Privacy (SP)* (pp. 770-787). IEEE. http://dx.doi.org/10.1109/SP54263.2024.00061

Ho, J., Jain, A., & Abbeel, P. (2020). Denoising diffusion probabilistic models. *Advances in neural information processing systems, 33*, 6840-6851. http://dl.acm.org/doi/abs/10.5555/3495724.3496298

Ho, J., & Salimans, T. (2021, December). Classifier-Free Diffusion Guidance. In *NeurIPS 2021 Workshop on Deep Generative Models and Downstream Applications*. https://openreview.net/forum?id=qw8AKxfYbI

Holmes, W., & Miao, F. (2023). *Guidance for generative AI in education and research*. UNESCO Publishing. https://doi.org/10.54675/EWZM9535

Hong, W., Ding, M., Zheng, W., Liu, X., & Tang, J. (2023). CogVideo: Large-scale Pretraining for Text-to-Video Generation via Transformers. In *The Eleventh International Conference on Learning Representations*. https://openreview.net/forum?id=rB6TpjAuSRy

Iqbal, N., Ahmed, H., & Azhar, K. A. (2022). Exploring teachers' attitudes towards using

ChatGPT. *Global Journal for Management and Administrative Sciences, 3*(4), 97-111.

http://dx.doi.org/10.46568/gjmas.v3i4.163

Iskender, A. (2023). Holy or unholy? Interview with open AI's ChatGPT. *European Journal of

Tourism Research, 34*, 3414-3424. https://doi.org/10.54055/ejtr.v34i.3169

Jhurree, V. (2005). Technology integration in education in developing countries: Guidelines to policy makers. *International Education Journal, 6*(4), 467-483.

https://www.learntechlib.org/p/103492/

KERIS. (2022). *2022 White paper on ICT in education in Korea*. Korea Education and Research

Information Service.

Khan, R. A., Jawaid, M., Khan, A. R., & Sajjad, M. (2023). ChatGPT-Reshaping medical

education and clinical management. *Pakistan Journal of Medical Sciences, 39*(2).

https://doi.org/10.12669/pjms.39.2.7653

Kling AI. (2024). *Kling*. https://klingai.com

Koehler, M., & Mishra, P. (2009). What is technological pedagogical content knowledge

(TPACK)?. *Contemporary issues in technology and teacher education, 9*(1), 60-70.

Kondratyuk, D., Yu, L., Gu, X., Lezama, J., Huang, J., Schindler, G., Hornung, R., Birodkar, V.,

Yan, J., Chiu, M.-C., Somandepalli, K., Akbari, H., Alon, Y., Cheng, Y., Dillon, J., Gupta,

A., Hahn, M., Hauth, A., Hendon, D., Martinez, A., ... & Jiang, L. (2023). Videopoet: A large language model for zero-shot video generation. *arXiv preprint arXiv:2312.14125*.

Kumar, L., Singh, D. K., & Ansari, M. A. (2024). Role of Video Content Generation in

Education Systems Using Generative AI. In *Integrating Generative AI in Education to*

*Achieve Sustainable Development Goals* (pp. 354-368). IGI Global.

http://dx.doi.org/10.4018/979-8-3693-2440-0.ch019

Lee, U., Han, A., Lee, J., Lee, E., Kim, J., Kim, H., & Lim, C. (2023). Prompt Aloud!:

Incorporating image-generative AI into STEAM class with learning analytics using prompt data. *Education and Information Technologies, 29*(8), 9575-9605.

http://dx.doi.org/10.1007/s10639-023-12150-4

Lee, U., Jeong, Y., Koh, J., Byun, G., Lee, Y., Lee, H., Eun, S., Moon, J., Lim, C., & Kim, H.

(2024a). I see you: teacher analytics with GPT-4 vision-powered observational assessment. *Smart Learning Environments, 11*(1), 48.

http://dx.doi.org/10.1186/s40561-024-00335-4

Lee, U., Jeon, M., Lee, Y., Byun, G., Son, Y., Shin, J., Ko, H., & Kim, H. (2024b).

LLaVA-docent: Instruction tuning with multimodal large language model to support art appreciation education. *Computers and Education: Artificial Intelligence, 7*, 100297.

http://dx.doi.org/10.1016/j.caeai.2024.100297

http://dx.doi.org/10.1016/j.caeai.2024.100297

Leiker, D., Gyllen, A. R., Eldesouky, I., & Cukurova, M. (2023, June). Generative AI for learning: investigating the potential of learning videos with synthetic virtual instructors.

In *International conference on artificial intelligence in education* (pp. 523-529). Cham:

Springer Nature Switzerland. https://doi.org/10.1007/978-3-031-36336-8_81

Lewis, P., Perez, E., Piktus, A., Petroni, F., Karpukhin, V., Goyal, N., Küttler, H., Lewis, M., Yih,

W., Rocktäschel, T., Riedel, S., & Kiela, D.Lewis, P., Perez, E., Piktus, A., Petroni, F.,

Karpukhin, V., Goyal, N., Küttler, H., Lewis, M., Yih, W., Rocktäschel, T., Riedel, S., &

Kiela, D.. (2020). Retrieval-augmented generation for knowledge-intensive nlp tasks. *Advances in Neural Information Processing Systems, 33*, 9459-9474.

Lim, W. M., Gunasekara, A., Pallant, J. L., Pallant, J. I., & Pechenkina, E. (2023). Generative AI and the future of education: Ragnarök or reformation? A paradoxical perspective from management educators. *The international journal of management education, 2*1(2), 100790. http://dx.doi.org/10.1016/j.ijme.2023.100790

Liu, Y., Cun, X., Liu, X., Wang, X., Zhang, Y., Chen, H., Liu, Y., Zeng, T., Chan, R. & Shan, Y. (2024a). Evalcrafter: Benchmarking and evaluating large video generation models. In *Proceedings of the IEEE/CVF Conference on Computer Vision and Pattern Recognition* (pp. 22139-22149). http://dx.doi.org/10.1109/CVPR52733.2024.02090

Liu, Y., Zhang, K., Li, Y., Yan, Z., Gao, C., Chen, R., Yuan, Z., Huang, Y., Sun, H., Gao, J., He, L., & Sun, L. (2024b). Sora: A review on background, technology, limitations, and opportunities of large vision models. *arXiv preprint arXiv:2402.17177*.

Lo, C. K. (2023). What is the impact of ChatGPT on education? A rapid review of the literature. *Education Sciences, 13*(4), 410. http://dx.doi.org/10.3390/educsci13040410

Luma AI. (2024). *Luma Dream Machine*. https://lumalabs.ai/dream-machine/creations

Marangunić, N., & Granić, A. (2015). Technology acceptance model: a literature review from 1986 to 2013. *Universal access in the information society*, *14*, 81-95. http://dx.doi.org/10.1007/s10209-014-0348-1

Marouf, M. A. A., Alchilibi, H., Sakr, D. K., Alkhayyat, A., & Pateriya, N. (2025). AI in Real-Time Student Performance Monitoring Using IoE. In *Role of Internet of Everything (IOE), VLSI Architecture, and AI in Real-Time Systems* (pp. 275-288). IGI Global Scientific Publishing. http://dx.doi.org/10.4018/979-8-3693-7367-5.ch019

Mazman Akar, S. G. (2019). Does it matter being innovative: Teachers' technology acceptance.

Education and Information Technologies, 24, 3415–3432.

https://doi.org/10.1007/s10639-019-09933-z

Memarian, B., & Doleck, T. (2023). ChatGPT in education: Methods, potentials and limitations.

Computers in Human Behavior: Artificial Humans, 100022.

http://dx.doi.org/10.1016/j.chbah.2023.100022

Mishra, P., Warr, M., & Islam, R. (2023). TPACK in the age of ChatGPT and Generative AI.

Journal of Digital Learning in Teacher Education, 39(4), 235-251.

https://doi.org/10.1080/21532974.2023.2247480

Mittal, U., Sai, S., Chamola, V., & Sangwan, D. (2024). A comprehensive review on generative

AI for education. IEEE Access, 12, 142733–142759.

https://doi.org/10.1109/ACCESS.2024.3468368

Mulyadi, D., Ali, M., Ropo, E., & Dewi, L. (2023). Correlational study: Teacher perceptions and the implementation of education for sustainable development competency for junior high school teachers. Journal of Education Technology, 7(2), 299-307.

http://dx.doi.org/10.23887/jet.v7i2.62728

Nazaretsky, T., Ariely, M., Cukurova, M., & Alexandron, G. (2022). Teachers' trust in

AI‑powered educational technology and a professional development program to improve it. British journal of educational technology, 53(4), 914-931.

https://doi.org/10.1111/bjet.13232

Noetel, M., Griffith, S., Delaney, O., Sanders, T., Parker, P., del Pozo Cruz, B., & Lonsdale, C.

(2021). Video improves learning in higher education: A systematic review. Review of

educational research, 91(2), 204-236. http://dx.doi.org/10.3102/0034654321990713

Norton, P., & Hathaway, D. (2010). Video production as an instructional strategy: Content

    learning and teacher practice. *Contemporary Issues in Technology and Teacher*

    *Education, 10*(1), 145-166.

Notari, M., Baumgartner, A., & Herzog, W. (2014). Social skills as predictors of communication,

    performance and quality of collaboration in project‑based learning. *Journal of Computer*

    *Assisted Learning, 30*(2), 132-147. http://dx.doi.org/10.1111/jcal.12026

Opara, E., Theresa, A. M. E., & Aduke, T. C. (2023). ChatGPT for teaching, learning and

    research: Prospects and challenges. *Glob Acad J Humanit Soc Sci, 5*.

OpenAI. (2022a, May 18). *DALL·E 2 research preview update*.

    https://openai.com/index/dall-e-2-update/

OpenAI. (2022b, November 30). *Introducing ChatGPT*. https://openai.com/index/chatgpt/

OpenAI. (2024). *Sora*. https://openai.com/sora/

Ousidhoum, N., Zhao, X., Fang, T., Song, Y., & Yeung, D. Y. (2021, August). Probing toxic

    content in large pre-trained language models. In *Proceedings of the 59th Annual Meeting*

    *of the Association for Computational Linguistics and the 11th International Joint*

    *Conference on Natural Language Processing* (Volume 1: Long Papers) (pp. 4262-4274).

    http://dx.doi.org/10.18653/v1/2021.acl-long.329

Owan, V. J., Abang, K. B., Idika, D. O., Etta, E. O., & Bassey, B. A. (2023). Exploring the

    potential of artificial intelligence tools in educational measurement and assessment.

    *Eurasia journal of mathematics, science and technology education, 19*(8), em2307.

    https://doi.org/10.29333/ejmste/13428

Owens, A. D., & Hite, R. L. (2022). Enhancing student communication competencies in STEM

    using virtual global collaboration project based learning. *Research in Science &*

*Technological Education, 40*(1), 76-102.

http://dx.doi.org/10.1080/02635143.2020.1778663

Peebles, W., & Xie, S. (2023). Scalable diffusion models with transformers. In *Proceedings of the IEEE/CVF International Conference on Computer Vision* (pp. 4195-4205).

http://dx.doi.org/10.1109/ICCV51070.2023.00387

Pellas, N. (2023). The influence of sociodemographic factors on students' attitudes toward AI-generated video content creation. *Smart Learning Environments, 10*(1), 57.

http://dx.doi.org/10.1186/s40561-023-00276-4

Premlatha, K. R., & Geetha, T. V. (2015). Learning content design and learner adaptation for adaptive e-learning environment: a survey. *Artificial Intelligence Review, 44*, 443-465.

http://dx.doi.org/10.1007/s10462-015-9432-z

Priestley, M., Edwards, R., Priestley, A., & Miller, K. (2012). Teacher agency in curriculum making: Agents of change and spaces for manoeuvre. *Curriculum inquiry, 42*(2), 191-214. http://dx.doi.org/10.1111/j.1467-873X.2012.00588.x

Qadir, J. (2022). Engineering education in the era of ChatGPT: Promise and pitfalls of generative AI for education. *2023 IEEE Global Engineering Education Conference (EDUCON)*, 1-9. https://doi.org/10.1109/EDUCON54358.2023.10125121

Ramadas, J. (2009). Visual and spatial modes in science learning. *International Journal of Science Education, 31*(3), 301-318. http://dx.doi.org/10.1080/09500690802595763

Ramesh, A., Pavlov, M., Goh, G., Gray, S., Voss, C., Radford, A., Chen, M., & Sutskever, I. (2021, July). Zero-shot text-to-image generation. In *International conference on machine learning* (pp. 8821-8831). https://proceedings.mlr.press/v139/ramesh21a.html

Runway AI. (2024). *Runway Gen-3*. https://runwayml.com/research/introducing-gen-3-alpha

Sánchez-Ruiz, L. M., Moll-López, S., Nuñez-Pérez, A., Moraño-Fernández, J. A., & Vega-Fleitas, E. (2023). ChatGPT challenges blended learning methodologies in engineering education: a case study in mathematics. *Applied Sciences, 13*(10), 6039-6060. https://doi.org/10.3390/app13106039

Shashidhara, V. V., Prabhu, R., & Chippar, P. (2024). Transforming ideas into products: Project based learning in prototyping, fabrication, and testing course for first year engineering students. *International Journal of Mechanical Engineering Education, 03064190241307026*. https://doi.org/10.1177/03064190241307026

Shen, C. (2024). Fair use, licensing, and authors' rights in the age of generative ai. *Northwestern Journal of Technology and Intellectual Property, 22*(1), 157-180.

Shepperd, P. (2024, September 3). Age restrictions and consent to use generative ai. *Artificial intelligence*. https://nationalcentreforai.jiscinvolve.org/wp/2024/09/03/age-restrictions-and-consent-to-use-generative-ai/?utm_source=chatgpt.com

Shetye, S. (2024). An evaluation of khanmigo, a generative ai tool, as a computer-assisted language learning app. *Studies in Applied Linguistics and TESOL, 24*(1). https://doi.org/10.52214/salt.v24i1.12869

Shi, L., Ding, A. C., & Choi, I. (2024). Investigating Teachers' Use of an AI-Enabled System and Their Perceptions of AI Integration in Science Classrooms: A Case Study. *Education Sciences, 14*(11), 1187. https://doi.org/10.3390/educsci14111187

Shu, D., Doss, C., Mondschein, J., Kopecky, D., Fitton-Kane, V., Bush, L., & Tucker, C. (2021, July). A pilot study investigating STEM learners' ability to decipher AI-generated video.

In 2021 ASEE Virtual Annual Conference. in China. *Applied Sciences, 14*(13), 5770. https://doi.org/10.18260/1-2--36601

Stumbrienė, D., Jevsikova, T., & Kontvainė, V. (2024). Key factors influencing teachers' motivation to transfer technology-enabled educational innovation. *Education and Information Technologies, 29*(2), 1697-1731. https://doi.org/10.1007/s10639-023-11891-6

Su, J., & Yang, W. (2023). Unlocking the power of ChatGPT: A framework for applying generative AI in education. *ECNU Review of Education, 6*(3), 355-366. https://doi.org/10.1177/20965311231168423

Sullivan, M., McAuley, M., Degiorgio, D., & McLaughlan, P. (2024). Improving students' generative AI literacy: A single workshop can improve confidence and understanding. *Journal of Applied Learning and Teaching, 7*(2). https://doi.org/10.37074/jalt.2024.7.2.7

Tippett, C. D. (2016). What recent research on diagrams suggests about learning with rather than learning from visual representations in science. *International Journal of Science Education, 38*(5), 725-746. https://doi.org/10.1080/09500693.2016.1158435

Tversky, A., & Kahneman, D. (1974). Judgment under uncertainty: Heuristics and biases: Biases in judgments reveal some heuristics of thinking under uncertainty. *Science, 185*(4157). 1124-1131. https://doi.org/10.1126/science.185.4157.1124

UNESCO (2023). *Guidance for generative AI in education and research*. https://doi.org/10.54675/EWZM9535

Upreti, K., Kushwah, V. S., Vats, P., Alam, M. S., Singhai, R., Jain, D., & Tiwari, A. (2024). A SWOT analysis of integrating cognitive and non‑cognitive learning strategies in education. *European Journal of Education*. https://doi.org/10.1111/ejed.12614

Van Dijk, J. (2005). *The deepening divide: Inequality in the information society*. SAGE.

   https://api.semanticscholar.org/CorpusID:153639791

Vaswani, A., Shazeer, N., Parmar, N., Uszkoreit, J., Jones, L., Gomez, A. N., Kaiser, &
   Polosukhun, I. (2017). Attention is all you need. *Advanced in Neural Information
   Processing Systems*, 30. https://dl.acm.org/doi/10.5555/3295222.3295349

Vayani, A., Dissanayake, D., Watawana, H., Ahsan, N., Sasikumar, N., Thawakar, O., Ademtew,
   H. B., Hmaiti, Y., Kumar, A., Kuckreja, K., Maslych, M., Ghallabi, W. A., Mihaylov, M.,
   Qin, C., Shaker, A. M., Zhang, M., Ihsani, M. K., Esplana, A., Gokani, M., … Khan, F. S.
   (2024). All languages matter: Evaluating LMMs on culturally diverse 100 languages.
   *arXiv preprint arXiv:2411.16508*. https://doi.org/10.48550/arXiv.2411.16508

Venkatesh, V., & Davis, F. D. (2000). A theoretical extension of the technology acceptance
   model: Four longitudinal field studies. *Management science, 46*(2), 186-204.

   https://doi.org/10.1287/mnsc.46.2.186.11926

Venkatesh, V., Morris, M. G., Davis, G. B., & Davis, F. D. (2003). User acceptance of
   information technology: Toward a unified view. *MIS Quarterly, 27*(3), 425-478.

   https://doi.org/10.2307/30036540

Viberg, O., Cukurova, M., Feldman-Maggor, Y., Alexandron, G., Shirai, S., Kanemune, S.,
   Wasson, B., Tømte, C., Spikol, D., Milrad, M., Coelho, R., & Kizilcec, R. F. (2024).
   What Explains Teachers' Trust in AI in Education Across Six Countries?. *International
   Journal of Artificial Intelligence in Education*, 1-29.

   https://doi.org/10.1007/s40593-024-00433-x

Villegas, R., Babaeizadeh, M., Kindermans, P. J., Moraldo, H., Zhang, H., Saffar, M. T., Castro,
   S., Kunze, J., & Erhan, D. (2023, Feb). Phenaki: Variable length video generation from open domain textual descriptions. In *International Conference on Learning Representations*. https://openreview.net/forum?id=vOEXS39nOF

Wang, W., Tu, Z., Chen, C., Yuan, Y., Huang, J. T., Jiao, W., & Lyu, M. (2024, August). All Languages Matter: On the Multilingual Safety of LLMs. In *Findings of the Association for Computational Linguistics ACL 2024* (pp. 5865-5877). https://aclanthology.org/2024.findings-acl.349.pdf

Wang, Y., Liu, C., & Tu, Y. F. (2021). Factors affecting the adoption of AI-based applications in higher education. *Educational Technology & Society, 24*(3), 116-129. https://www.jstor.org/stable/27032860

Weerakoon, O., Leppänen, V., & Mäkilä, T. (2024, June). Enhancing Pedagogy with Generative AI: Video Production from Course Descriptions. In *Proceedings of the International Conference on Computer Systems and Technologies 2024* (pp. 249-255). https://doi.org/10.1145/3674912.3674922

Trust, T., Whalen, J., & Mouza, C. (2023). ChatGPT: challenges, opportunities, and implications for teacher education. *Contemporary Issues in Technology and Teacher Education, 23*(1), 1-23. https://www.learntechlib.org/primary/p/222408/

Woolfitt, Z. (2015). The effective use of video in higher education. *Lectoraat Teaching, Learning and Technology Inholland University of Applied Sciences, 1*(1), 1-49. https://www.inholland.nl/media/10230/the-effective-use-of-video-in-higher-education-woolfitt-october-2015.pdf

Wozney, L., Venkatesh, V., & Abrami, P. (2006). Implementing computer technologies: Teachers' perceptions and practices. *Journal of Technology and teacher education, 14*(1), 173-207. https://api.semanticscholar.org/CorpusID:152558814

Yu, T., Yang, W., Xu, J., & Pan, Y. (2024). Barriers to industry adoption of AI video generation tools: A study based on the perspectives of video production professionals in China. *Applied Sciences, 14*(13), 5770. https://doi.org/10.3390/app14135770

Yue, M., Jong, M. S. Y., & Ng, D. T. K. (2024). Understanding K–12 teachers' technological pedagogical content knowledge readiness and attitudes toward artificial intelligence education. *Education and Information Technologies*, 1-32. https://doi.org/10.1007/s10639-024-12621-2

Zhai, X. (2024). Transforming teachers' roles and agencies in the era of generative ai: Perceptions, acceptance, knowledge, and practices. *Journal of Science Education and Technology*, 1-11. https://doi.org/10.1007/s10956-024-10174-0

Zhang, D., Zhou, L., Briggs, R. O., & Nunamaker Jr, J. F. (2006). Instructional video in e-learning: Assessing the impact of interactive video on learning effectiveness. *Information & management, 43*(1), 15-27. https://doi.org/10.1016/j.im.2005.01.004

Zhang, L., Rao, A., & Agrawala, M. (2023). Adding conditional control to text-to-image diffusion models. In *Proceedings of the IEEE/CVF International Conference on Computer Vision* (pp. 3836-3847). http://doi.org/10.1109/ICCV51070.2023.00355

Zhang, Q., Ye, Y., Liu, S., Ma, H., Zheng, L., Ren, Y., & SUO, L. (2022). Factors affecting the willingness of Chinese users to continue using online education platforms in Yunnan. *Human Sciences, 14*(2), 137-159.

Zhao, S., Chen, D., Chen, Y. C., Bao, J., Hao, S., Yuan, L., & Wong, K. Y. K. (2024). Uni-controlnet: All-in-one control to text-to-image diffusion models. *Advances in Neural Information Processing Systems, 36.* https://arxiv.org/abs/2305.16322

Zhao, Y., & Frank, K. A. (2003). Factors affecting technology uses in schools: An ecological perspective. *American Educational Research journal, 40*(4), 807-840.

https://doi.org/10.3102/00028312040004807

# Appendix

## Appendix 1 - Interview questionnaire

*Interview questionnaire for research*

| Elements | Questions |
|---|---|
| Basic information | <ul><li>Could you tell me your name, gender, nationality, and age?</li><li>What is your major, occupation, and experience in this profession?</li></ul> |
| TPACK & TAM | <ul><li>What was your impression when you first experienced the video generation model?<ul><li>○ Which features of the video generation model did you find most interesting or useful?</li><li>○ What technical challenges did you encounter while using the video generation model?</li><li>○ How easy (or difficult) did you find using the video generation model?</li></ul></li><li>Do you have knowledge about various technological tools and platforms related to the video generation model?</li><li>What types of learning content do you think can be delivered using the video generation model?<ul><li>○ In your opinion, how can the video generation model assist in your lessons? (Examples for better understanding: engagement, comprehension, etc.)</li><li>○ What teaching methods do you think are appropriate to effectively utilize the video generation model in lessons?</li></ul></li></ul> |

- What strategies do you consider suitable when planning lessons with the video generation model?

- How do you think the video generation model should be integrated with other educational technologies?

- How do you think the interaction between technology, pedagogy, and content should work in lessons using the video generation model?

- What ethical issues do you foresee when using the video generation model in an educational context?

- What institutional support do you think is needed to use the video generation model in education?

- How useful do you think the video generation model is in enhancing the effectiveness of lessons?

- What is your opinion about utilizing the video generation model in education?

- Would you be willing to use the video generation model in your lessons?
  - If you plan to introduce the video generation model into your lessons, how and how often do you intend to use it?

| | |
|---|---|
| Others | - Do you have any additional opinions or comments? |

**Appendix 2 - Educational Guideline**

*This is a translated version of an educational guideline. The original guideline is written in Korean.*

User Guide

- Login

  - Access the link and log in (Dream Machine).

    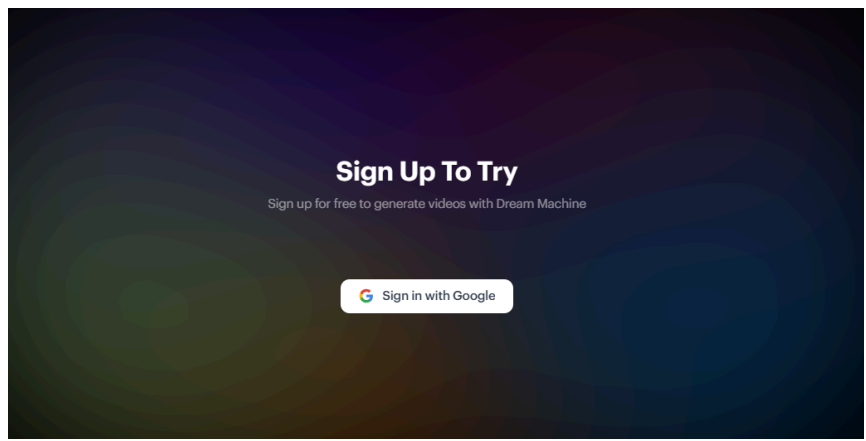

  - Enter your desired command (prompt) in the input box in the center and press Enter or click the upward arrow on the right. You can insert a source image by clicking the "Image Display" button on the left.

    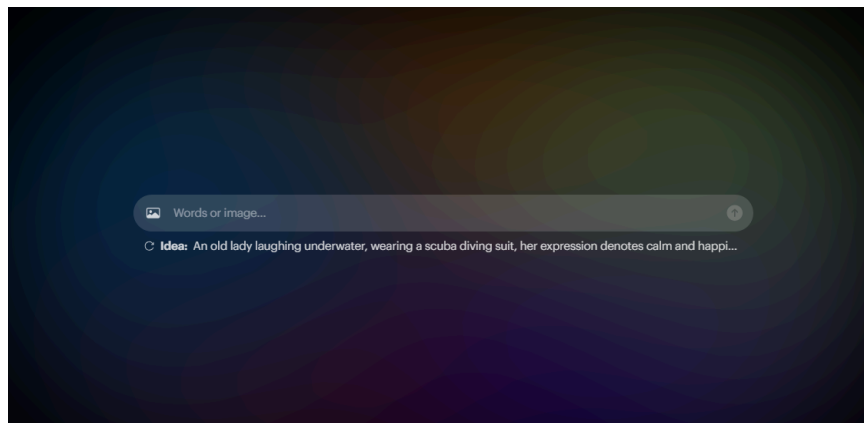

- (Example) Insert the image of a dog below and type "I want to make this dog jump around excitedly." Then click the arrow. The video in progress will appear in the queue below.

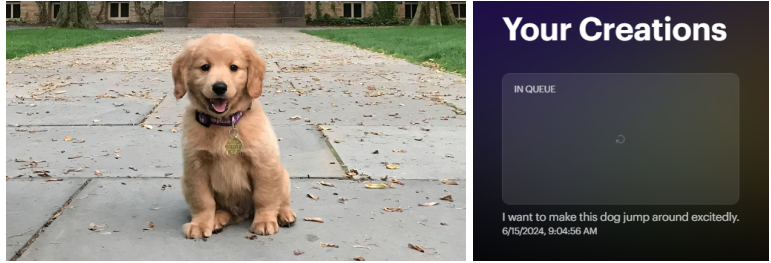

- Wait approximately two minutes for the video to complete (the description changes from "IN QUEUE" to "DREAMING..." before disappearing). Click the video to play it.

- Click "DIRECT LINK" below and then click "..." in the bottom-right corner to find the "DOWNLOAD" button to save the completed video.

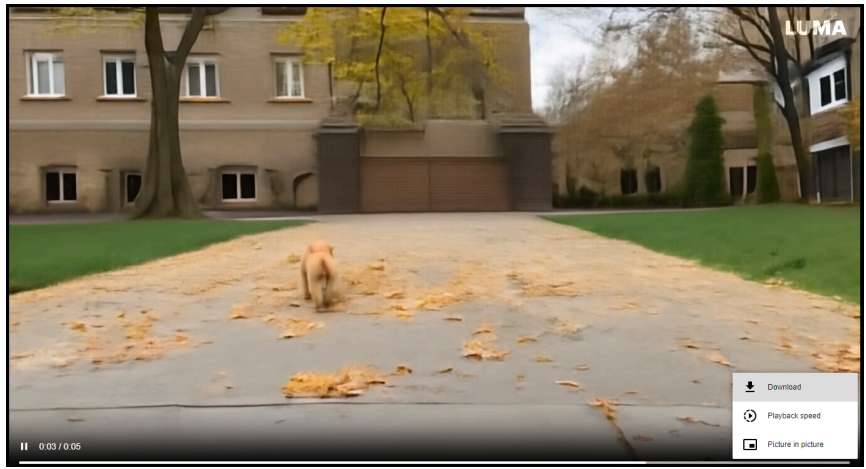

- Notes
  - If multiple users are accessing the service simultaneously, you may need to wait your turn in the queue. Even if the video does not process immediately, it will be completed eventually.
- Expected Applications

- - English
    - Teaching Context
      - During textbook lessons, create videos for parts of the text to engage students. Continuous video production allows for the creation of a series in the form of a movie.
    - Example Prompt
      - However, as soon as Joe and Simon had reached the summit, Simon saw another snowstorm coming from the east. Moreover, storm clouds were quickly approaching the north ridge as the two men started heading down the mountain. The climbers, who were not used to dealing with snowstorms, decided to descend in zero visibility.
      - (From High School English I, Unit 6, "When the Going Gets Tough")
  - Example Video

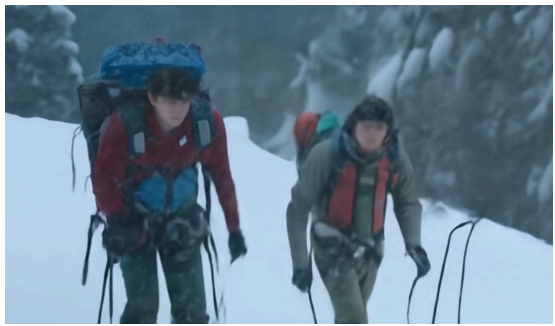

- Science
  - Teaching Context
    - Convert written explanations of abstract microscopic concepts into video explanations to improve students' understanding.

- Example Prompt

  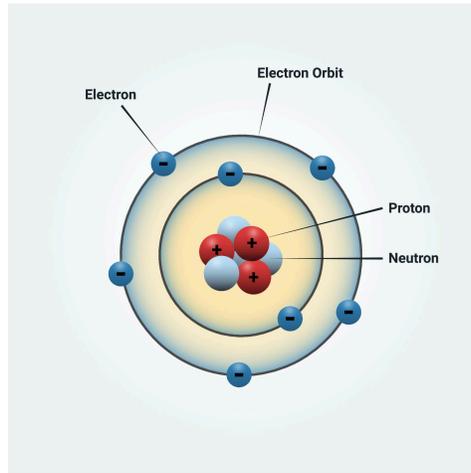

  - An atom has a central nucleus with positively charged protons and neutral neutrons. Negatively charged electrons orbit the nucleus. The number of protons equals the number of electrons, balancing the atom's charge. This structure is the basic building block of all matter.
- [Example Video](#)

  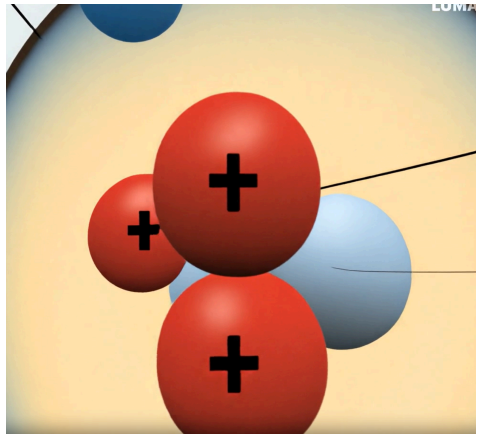

- Art

  - Teaching Context

- Animation Production Assignment: Use custom-created images and animate them, demonstrating camera movements and scene transitions as a reference.

- Video Art Creation: Animate famous paintings, turning them into motion art.

- Understanding Filming Techniques: Demonstrate differences between high-angle and low-angle shots.

- Example Prompts

  - Let the light in this picture sparkle and make it look like a moving animation.

  - Create a video of the woman in the painting looking forward. (Gerhard Richter's Betty)

    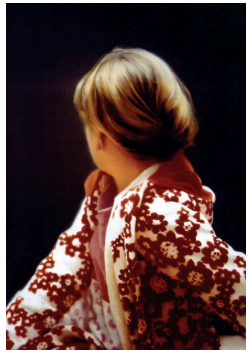

- Example Videos

  - [Art Example 1 (Scene Transition Reference)](#)

  - [Art Example 2 (Transformed Masterpiece)](#)

  - [Art Example 3-1 (Same Subject, Different Angles)](#)

  - [Art Example 3-2](#)

  - [Art Example 3-3](#)

- Social Studies
  - Teaching Context
    - Represent complex economic concepts like inflation and deflation through motion graphics.
    - Visualize historical events like the progression of wars using motion graphics.
  - Example Prompt
    - Motion graphic explaining inflation and deflation
  - Example Video
    - [Social Studies Example](#)
- Korean Language
  - Teaching Context
    - Recreate scenes from literary works to deepen understanding.
    - Create videos for discussion topics.
  - Example Prompt
    - In the small house where five family members live together, a forced eviction notice arrives. Upon hearing the sound of the eviction notice, Father seems to have stopped his work and returned, and I followed beside him as he inspected the toolbox containing his tools. Mother, aware that trouble could ensue later if she didn't quickly remove the aluminum plate engraved with the unauthorized building number, was using a kitchen knife to pry out the nails holding the plate *"The Dwarf"*.

- For discussion topics: Factories emitting smoke, then transitioning to cleaner factories and renewable energy sources.
  - Example Videos
    - [Korean Language Arts Example 1](#)
    - [Korean Language Arts Example 2](#)
- Advanced Applications
  - Detailed Descriptions of Characters/Objects and Emotions
    - A joyful young woman with auburn hair in a blue sundress stands.
    - An elderly man with silver hair, dressed in a brown coat.
  - Detailed Descriptions of Settings and Lighting
    - Under the soft, diffused light of an overcast sky.
    - A dusty attic, illuminated by soft, golden rays of sunlight filtering through a small, grimy window.
- Camera Angles and Focus Descriptions
  - Low-angle / High-angle / Wide-angle / Side-angle / Close-up, etc.
  - Sharp focus / Soft focus / Out of focus / Selective focus, etc.

**Appendix 3 - Code Book**

| Theme | Subtheme | Code |
|-------|----------|------|
| Tool perception | Impression | Positive impression |
| | | Negative impression |
| | Usage experience | Ease of use |
| | | Usefulness |
| | | Experience with similar tools |
| | Willingness to use | Willingness to use |
| Instructional design and implementation | Design of instructional content and methods | Instructional design |
| | | Learning design |
| | | Teaching strategies and pedagogy |
| | | Assessment methods |
| | | Integration with other tools |
| | | Literacy education |
| | Expected effects | Understanding of subject content |
| | | Improvement in literacy (AI, media, digital) |
| | | Increased engagement and interest |
| | | Improved efficiency/effectiveness in lesson preparation |
| | | Enhancement of creativity and expressiveness |
| | | Provision of personalized materials |
| | Potential issues | Data bias |
| | | Neglect and violation of rights |
| | | Misuse of tools |
| | | Widening learning gaps |
| | | Decline in learning motivation |
| | | Misconceptions |
| | | Harmful outputs |
| | | Lack of consistency |
| | Teacher competency | Technical competency of teachers |
| | | Competency in lesson design and management |
| Tool improvement and support | Policy improvements | Ensuring teacher autonomy and authority |
| | | Devices and infrastructure |

| | | Administrative support |
|---|---|---|
| | | User training and awareness improvement |
| | | Improvement of model performance |
| | Technical improvements | Enhancement of service usability |
| | | Enhancement of educational usefulness and safety |